\documentclass[11pt,a4paper]{article}
\pdfoutput=1
\usepackage{jheppub}
\usepackage[utf8]{inputenc}

\usepackage{color}
\usepackage{amsmath}
\usepackage{mathtools}
\usepackage{esint}
\usepackage{pifont}
\usepackage{bbold}

\usepackage{bbm}
\usepackage{verbatim}   
\usepackage{subfigure}
\usepackage{acronym}

\usepackage{amsfonts}
\usepackage{amssymb}
\usepackage{mathrsfs}
\usepackage{graphicx}
\usepackage{multirow}
 \usepackage{slashed}

\usepackage{caption}

\usepackage{soul}

\definecolor{mypurple}{RGB}{164,64,214}

\title{Properties of Discrete Black Hole Hair}

\author{Isabel Garcia Garcia}
\emailAdd{isabel@kitp.ucsb.edu}

\affiliation{Kavli Institute for Theoretical Physics, University of California, Santa Barbara, CA 93106, USA}

\abstract{We revisit the physical effects of discrete $\mathbb{Z}_p$ gauge charge on black hole thermodynamics, building on the seminal work of Coleman, Preskill, and Wilczek.
Realising the discrete theory from the spontaneous breaking of an Abelian gauge theory, we consider the two limiting cases of interest, depending on whether the Compton wavelength of the massive vector is much smaller or much larger than the size of the black hole -- the so-called thin- and thick-string limits respectively.
We find that the qualitative effect of discrete hair on the mass-temperature relationship is the same in both regimes, and similar to that of unbroken $U(1)$ charge: namely, a black hole carrying discrete gauge charge is always colder than its uncharged counterpart.
In the thick-string limit, our conclusions bring into question some of the results of Coleman et al., as we discuss.
Further, by considering the system to be enclosed within a finite cavity, we argue how the unbroken limit may be smoothly defined, and the unscreened electric field of the standard Reissner-Nordstr\"om solution recovered.}

\begin{document}

\maketitle


\section{Introduction}
\label{sec:intro}

At the classical level, no-hair theorems restrict the properties of a black hole (BH) that can be measured by an asymptotic observer to mass, charge, and angular momentum.
Quantum mechanically, this restriction no longer applies, allowing BHs to carry a wide variety of internal quantum numbers.
An example of such ``quantum hair'' is charge under a discrete gauge symmetry. 

Discrete gauge symmetries can arise from spontaneous symmetry breaking of continuous ones if the charge of the condensate responsible for the breaking is larger than the charge quantum.
The simplest example is the breaking of a $U(1)$ gauge symmetry down to a local $\mathbb{Z}_p$ as a result of the non-zero vacuum expectation value (vev) $v$ of a Higgs field carrying $U(1)$ charge $g p$ (with $p > 1$).
Under the leftover discrete symmetry the Higgs field transforms trivially, but a state $\eta$ with unit charge $g$ under the original $U(1)$ would transform as
\begin{equation}
	\eta \rightarrow e^{i\frac{2 \pi}{p}} \eta \ .
\label{eq:ABphase}
\end{equation}

In the broken phase, both Higgs and vector fields become massive, and must be integrated out in order to obtain a low energy effective description of the theory at scales $\ll v$.
In the effective field theory realm, the only effect of the discrete symmetry is to forbid ${\mathbb Z}_p$-violating interactions among light degrees of freedom, just as would be the case if the symmetry were global.
At energies of order $\sim v$, the massive scalar and vector states can be excited, obviously differentiating the gauge and global theories.
Moreover, the local theory also contains solitonic degrees of freedom:
flux-tubes, or cosmic strings, inside of which magnetic flux of the broken $U(1)$ remains confined \cite{Nielsen:1973cs}.
The amount of magnetic flux carried by cosmic strings is quantized in units of $2 \pi / (g p)$.

When $p>1$, and there is truly a non-trivial discrete symmetry left unbroken, the Higgs mechanism is only effective at screening charge modulo $p$, with states carrying smaller electric charge still subject to long-range interactions.
Specifically, interactions between cosmic strings and particles charged under the leftover discrete symmetry are dominated at low energies by Aharonov-Bohm scattering \cite{Aharonov:1959fk, Alford:1988sj}.
(Notice the phase in Eq.(\ref{eq:ABphase}) is just the Aharonov-Bohm phase that a particle with unit charge picks up when looped around a flux-tube carrying one unit of magnetic flux.)
By means of Aharonov-Bohm scattering experiments, discrete electric charge becomes an asymptotic observable, crucially differentiating the local and global versions of the theory.

The local nature of the discrete symmetry allows for a version of Gauss' law to be constructed, relating the total ${\mathbb Z}_p$ charge inside some volume to
an operator defined on an asymptotic bounding surface \cite{Alford:1989ch,Preskill:1990bm}.
In turn, this ensures that the total discrete charge inside any closed surface cannot be changed by any local process taking place within the interior region -- a category which presumably includes collapse to a BH.
As a result, if some initial distribution of matter carries non-zero discrete charge, the BH at the end of the collapse process will retain this discrete hair.
Notice this type of hair is quantum in nature, as the existence of a leftover discrete symmetry after spontaneous symmetry breaking depends crucially on quantization of charge,
as well as on the associated Aharonov-Bohm scattering,
and is therefore outside the scope of applicability of classical no-hair theorems.

It was first pointed out that BHs can carry discrete gauge charge as quantum hair in the context of Abelian gauge theories in \cite{Krauss:1988zc},
with \cite{Alford:1989ch,Alford:1990pt} extending the argument to the non-Abelian case. (Other types of quantum hair were discussed in \cite{Bowick:1988xh}.)
Further, Coleman, Preskill, and Wilczek, in a beautiful paper \cite{CPW}, studied the physical effects of Abelian discrete hair on BH thermodynamics,
with an emphasis on how the relationship between mass and temperature is altered.
In the context of a discrete gauge theory realised through a model of spontaneous symmetry breaking,
the authors of \cite{CPW} argued that the leading effect of local ${\mathbb Z}_p$ charge is due to a virtual process in which a cosmic string is nucleated on the horizon surface, grows to envelop the BH,
and re-annihilates at the antipodal point.
By extending the traditional Euclidean path integral approach to BH thermodynamics, they showed that this virtual process is expressed in the Euclidean formalism through the existence of instantons, specifically vortex solutions in the $r-\tau$ plane ($\tau$ being the periodically identified Euclidean time coordinate).
The typical Euclidean vortex radius corresponds in the space-time picture to the thickness of the nucleated string, and there are two limiting cases of interest depending on whether the string thickness is much smaller or much larger than the size of the BH. These two cases correspond to the Compton wavelength of the massive vector being much smaller or much larger than the horizon radius $r_+$, and are realised in the limits $v r_+ \gg 1$ and $v r_+ \ll 1$ respectively.
\footnote{Strictly speaking, the thin- and thick-string scenarios correspond to $m_v r_+ \gg 1$ and $m_v r_+ \ll 1$, with $m_v = g p  v $ the vector mass.
However, we assume everywhere in this work that $g p = \mathcal{O}(1)$, and use $m_v \sim v$.}

The space-time interpretation put forward in \cite{CPW} is further reinforced by considerations regarding the electric field that an asymptotic observer can measure. 
Unlike the situation for unbroken $U(1)$ charge, there is no classical electric field for screened electric charge
-- at the classical level, where there is no relevant notion of charge quantization, the Higgs mechanism screens \emph{all} of the charge.
However, a non-zero expectation value of the electric field in the radial direction is present whenever the BH carries discrete charge $Q$ that is not an exact multiple of $g p$.
Parametrically, it is given by \cite{CPW}
\begin{equation}
	\langle E_r (r) \rangle \sim \sin \left( \frac{2 \pi Q}{g p} \right) F_{r \tau} (r) e^{- \Delta S_v} \ ,
\label{eq:Ervortex}
\end{equation}
where $F_{r \tau} (r)$ corresponds to the Euclidean magnetic field component for the unit-flux vortex, and $\Delta S_v$ is the value of the Euclidean action of the vortex solution.
The sine factor ensures that this expectation value vanishes whenever $Q/g$ is a multiple of $p$, in which case the Higgs mechanism provides perfect screening also at the quantum level.
Although the origin of Eq.(\ref{eq:Ervortex}) will become clearer later, its qualitative form is consistent with the previous discussion:
as a cosmic string is nucleated on the horizon surface, and grows to envelop the BH, the moving magnetic flux that is confined inside the string generates an electric field in the radial direction.
This electric field is only non-zero when a string is nucleated, and therefore its expectation value should be suppressed by a factor related to the corresponding tunnelling rate.
Such suppression is provided by the last factor in Eq.(\ref{eq:Ervortex}).
Expectation values for higher powers of the electric field are found to feature the same exponential suppression $\sim e^{- \Delta S_v}$ \cite{CPW}, providing unambiguous evidence of the non-perturbative nature of this effect.

In the thin-string limit, the results of \cite{CPW} concerning the thermodynamic properties of BHs carrying discrete hair are consistent with expectations:
in this regime, a BH that carries screened electric charge is found to be \emph{colder} than an uncharged BH of the same mass.
The size of the correction is exponentially suppressed, but qualitatively the effect is the same as for unscreened hair. 
\footnote{See also \cite{Dowker:1991qe} for further details relevant to the thin-string regime.}
On the other hand, in the thick-string case, the authors of \cite{CPW} claim that the sign of the correction to the BH mass-temperature relationship flips: now, a BH carrying discrete charge is \emph{hotter} than its uncharged counterpart.
As for the possibility of taking the limit $v \rightarrow 0$ that would smoothly interpolate between screened and unscreened hair, the authors of \cite{CPW} note that such limit cannot be taken, within the framework of their calculation, while maintaining the validity of the semiclassical expansion.

We believe that the results found in \cite{CPW} in the thick-string limit beg for further explanation on two grounds.
First, that the correction to the mass-temperature relationship has opposite sign in the thin- and thick-string limits requires that as $v$ is decreased from $v r_+ \gg 1$ to $v r_+ \ll 1$ there is some intermediate value of the symmetry breaking scale at which the effect of screened hair on the thermodynamic properties of a BH somehow vanishes.
Second, one would expect that smoothly decreasing $v$ from the thin- to the thick-string regime would somehow resemble the transition from a Schwarzschild BH (formally $v r_+ \rightarrow \infty$) to a BH carrying unscreened charge.
That the correction to the BH mass-temperature relationship in the thick-string limit is opposite to that of standard electric charge seems contrary to this expectation.

In this work, we revisit the calculations of \cite{CPW}, focusing on the regime in which gravitational back-reaction can be neglected, and with a particular emphasis on the thick-string limit $v r_+ \ll 1$.
Although we are in agreement regarding the overall size of the effect of discrete charge in this limit, we find that the sign of the mass-temperature correction is in fact the same as in the thin-string case.
That is, we find that both in the thin- and thick-string regimes, a BH carrying discrete charge is \emph{colder} than an uncharged BH of the same mass, with the size of the correction getting larger as $v$ decreases.
Further, by considering the BH to be enclosed within a cavity of finite size, we show how the $v \rightarrow 0$ limit may be smoothly taken, and the standard properties of unscreened hair recovered.
Our calculation shows how a classical component of the radial electric field emerges, and although it does not yet allow for the full $v \rightarrow 0$ limit to be realised, it provides non-trivial evidence as to how a smooth transition may be well defined within finite-sized systems.

This article is organized as follows. 
In section~\ref{sec:pathintegral} we review some salient features of the Euclidean path integral approach to BH thermodynamics, especially the definition of the grand-canonical partition function, and the subsequent projection onto a sector of fixed charge.
Section~\ref{sec:discreteInfiniteBox} focuses on the effect of screened electric charge in the context of infinite, asymptotically flat space. After reviewing the thin-string treatment of \cite{CPW}, we present our thick-string results in \ref{sec:thick}, and elaborate on how our results differ from those of Coleman et al.
In section~\ref{sec:discreteFiniteBox}, we refine the previous thick-string analysis by considering the BH to be enclosed within a finite spherical cavity, and discuss how considering a finite-sized system may allow for a smooth $v \rightarrow 0$ limit. Finally, section~\ref{sec:conclusions} contains a brief summary of our conclusions.
Everywhere in this paper, we set $\hbar = c = k_B = 1$, but keep factors of Newton's constant $G$.

\section{Path integral approach to black hole thermodynamics}
\label{sec:pathintegral}

In quantum field theory the Euclidean path integral is an object of the form:
\begin{equation}
	Z_{\rm QFT} = \int d [\varphi] e^{- S_E [\varphi]} \ ,
\label{eq:QFTZ}
\end{equation}
where $d [\varphi]$ represents a measure in the space of matter fields, and the sum must include all field configurations satisfying some given boundary conditions.
The meaning of this expression will crucially depend on what these boundary conditions are.
On a flat background, imposing periodicity in the Euclidean time coordinate $\tau$ with period $\beta$,
the path integral takes a special meaning: it corresponds to the partition function appropriate to describe the statistical mechanics of a thermodynamic
ensemble at temperature $T = \beta^{-1}$.

In the context of quantum gravity, Eq.(\ref{eq:QFTZ}) needs to be extended to include a sum over metrics.
By analogy with the results in quantum field theory, one would expect that by imposing periodicity in $\tau$,
the gravitational path integral would provide information about the statistical mechanics of gravitational systems.
Although this procedure is far from being rigorously established, this basic idea has proved valuable in studying the thermodynamics of self-gravitating systems such as BHs.

Our ignorance regarding how to compute the gravitational path integral obliges us to resort to approximations.
The best one can do is expand around field configurations that extremize the Euclidean action:
\begin{equation}
	Z_{\rm QG} = \int d [g] d [\varphi] e^{- S_E [g, \varphi]} \simeq e^{- S_E [\hat g, \hat \varphi]} \ ,
\label{eq:QGZapprox}
\end{equation}
where $\hat g$ and $\hat \varphi$ refer to the classical solutions to the Euclidean equations of motion with the appropriate boundary conditions.
The right-hand-side of Eq.(\ref{eq:QGZapprox}) then corresponds to the leading order term in the semiclassical evaluation of the gravitational path integral.

The sum over metrics in Eq.(\ref{eq:QGZapprox}) divides into topologically distinct sectors, all of which may contribute to leading order in the semiclassical expansion.
If we are only interested in the properties of a certain topological sector, we can further restrict the sum over metrics to those with the appropriate topology.
In \cite{Gibbons:1976ue}, Gibbons and Hawking argued that the thermodynamics of a BH with fixed temperature can be obtained from the gravitational path integral by demanding
periodicity in $\tau$, and further restricting the sum over metrics to those with $\mathbb{R}^2 \times S^2$ topology.

In this section, we review how this formalism is applied to study the thermodynamics of BHs within a canonical ensemble (fixed temperature and system size).
We discuss the Schwarzschild case in \ref{sec:SchwBH}, and in \ref{sec:RNBH} we extend the discussion to BHs carrying Abelian gauge charge.
The purpose of this section is not to provide a comprehensive review of the Eculidean path integral approach to BH thermodynamics, but rather to highlight the main concepts
that will later be relevant.

\subsection{Schwarzschild black hole}
\label{sec:SchwBH}

The Schwarzschild solution to the Euclidean Einstein's equations is given by
\begin{equation}
	d s^2 = g_{\tau\tau} (r) d\tau^2 + g_{rr} (r) dr^2 + r^2 d\Omega^2 \ , \qquad {\rm with} \qquad g_{\tau\tau} (r) = g_{rr} (r)^{-1} = 1 - \frac{r_+}{r} \ ,
\label{eq:SchwMetric}
\end{equation}
and it is just the analytic continuation of that part of the Lorentzian metric lying outside or at the event horizon.
Demanding the absence of a conical singularity at $r=r_+$ requires the Euclidean time coordinate $\tau$ be cyclic with period $\tilde \beta = 4 \pi r_+$.
A Euclidean BH is therefore topologically ${\mathbb R}^2 \times S^2$ -- the ${\mathbb R}^2$ corresponding to the $r - \tau$ plane, centred at the event horizon.
Contrary to their Lorentzian counterparts, Euclidean BHs are completely smooth solutions, featuring neither an interior nor a singularity.

Studying the thermodynamic properties of a Schwarzschild BH within a canonical description requires fixing both the size of the system as well as its temperature.
As discussed in \cite{York:1986it}, we can fix the system size by considering the Schwarzschild solution within a finite region $r_+ \leq r \leq r_B$.
This corresponds to a BH at the centre of a spherical cavity, with the size of the system given by the area $4 \pi r_B^2$ of the cavity wall.
We define the temperature to be $T = \beta^{-1}$ at the wall itself, which imposes the relationship $\beta = 4 \pi r_+ \sqrt{g_{\tau \tau} (r_B)}$.
\footnote{This is just the standard expression for the (inverse) temperature of a BH of size $r_+$ in asymptotically flat space corrected by the appropriate redshift factor.
The equation $\beta  =  4 \pi r_+ \sqrt{g_{\tau \tau} (r_B)}$ implicitly defines $r_+$ as a function of $\beta$ and $r_B$, and has two physical solutions when $3 \sqrt{3} \beta \leq 8 \pi r_B$ \cite{York:1986it}.
In this paper, we will only be concerned with the solution that survives in the limit $r_B \rightarrow \infty$.
In this case, $r_+ \sim \beta$, and an expansion in $r_+ / r_B \sim \beta / r_B \ll 1$ is possible.}
In a system of finite size, the relationship between the mass and temperature of a BH reads
\begin{equation}
	M (\beta, r_B) \equiv \frac{r_+}{2G} \simeq \frac{\beta}{8 \pi G} \left( 1 + \frac{\beta}{8 \pi r_B} \right) \ ,
\label{eq:betarp}
\end{equation}
where the right-hand-side only includes the leading $r_+ / r_B$ correction.

In the absence of matter fields, the only contribution to the Euclidean action comes from the gravitational sector, and can be written as
\begin{equation}
	S_{g} = - \frac{1}{16 \pi G} \int_{\mathcal{M}} d^4 x \sqrt{g} R - \frac{1}{8 \pi G} \int_{\partial \mathcal{M}} d^3 x \sqrt{h} K \ ,
\label{eq:SESchw}
\end{equation}
where $\mathcal{M}$ refers to the region $r \leq r_B$, $\partial \mathcal{M}$ is the boundary at $r=r_B$,
and $h_{\mu \nu}$ and $K$ correspond to the induced metric and extrinsic curvature of the boundary respectively.
The second term in Eq.(\ref{eq:SESchw}) corresponds to the Gibbons-Hawking-York boundary term \cite{York:1972sj,Gibbons:1976ue,York1986},
whose role is to ensure that the action is stationary under variations of the metric around backgrounds that satisfy Einstein's equations.
In the absence of a traceful energy-momentum tensor, $R$ vanishes, and the only contribution to the action comes from this second term.
An intrinsic ambiguity in the definition of Eq.(\ref{eq:SESchw}) stems from the fact that adding any function that depends only on the boundary data does not change the equations of motion \cite{Brown:1992br,Brown:1992bq}.
The standard approach is to require that the action vanishes on some reference background \cite{Hawking:1995fd}, which is conveniently chosen to be flat space.
Such a choice matches the asymptotic behaviour of the Schwarzschild solution, and leads to a finite action in the $r_B \rightarrow \infty$ limit.
With this prescription, the appropriately regularised Schwarzschild action reads
\begin{equation}
	S_g (\beta, r_B) = - \frac{\beta}{2 G} \left. \partial_r \left( r^2 (\sqrt{g_{\tau \tau} (r)}  -  1) \right) \right|_{r_B} \simeq \frac{\beta^2}{16 \pi G} \left( 1 + \frac{\beta}{8 \pi r_B} \right) \ . 
\label{eq:deltaSBH}
\end{equation}

To leading order in the semiclassical expansion the free energy of the ensemble is therefore given by $F(\beta, r_B) \simeq S_g (\beta, r_B) / \beta$,
and all other thermodynamic quantities can be obtained by differentiation.
For instance, the internal energy of the system can be written as
\begin{equation}
 	U (\beta, r_B) = \left( \frac{\partial (\beta F)}{\partial \beta} \right)_{r_B} \simeq
		\frac{\beta}{8 \pi G} \left( 1 + \frac{\beta}{8 \pi r_B} \right) + \frac{G}{2 r_B} \left( \frac{\beta}{8 \pi G} \right)^2 \ ,
\label{eq:SchwU}
\end{equation}
where the first term is the mass of the BH, as given in Eq.(\ref{eq:betarp}), and the second corresponds to minus the gravitational self-energy of the system.
Only in the limit $r_B \rightarrow \infty$ mass and internal energy coincide, leading to the familiar expression $M (\beta) = U (\beta) = \beta / (8 \pi G)$.

\subsection{Charged black hole}
\label{sec:RNBH}

In this section, we extend the previous discussion to the case of BHs that carry charge under an Abelian gauge symmetry.
Now, boundary conditions for the gauge fields must be specified, and there are two choices: either fixing the boundary value of the one-form potential $A$,
or that of the two-form field-strength $F$. The former choice corresponds to a system with fixed chemical potential, whereas the latter is appropriate for fixing the total electromagnetic charge \cite{Hawking:1995ap}.
The system will then be described in the context of the grand-canonical and canonical ensembles respectively.

Here we will be occupied with the case of a BH with fixed electric charge $Q$, but to obtain the relevant partition function
we will follow the alternative procedure described in \cite{CPW}, which consists on imposing boundary conditions on $A$, and subsequently projecting the system onto a state of given
charge by introducing the appropriate projection operator into the path integral.
For a $U(1)$ gauge theory, this procedure is uncontroversially equivalent to fixing the boundary value of $F$, as discussed thoroughly in \cite{Hawking:1995ap},
but it can also be extended to the case of discrete gauge charge, which is the main topic of this paper.

The static, spherically symmetric solution to the Einstein-Maxwell equations with ${\mathbb R}^2 \times S^2$ topology and non-zero electromagnetic charge is the Euclidean Reissner-Nordstr\"om solution.
However, in the region far from extremality, $G Q^2 / r_+^2 \ll 1$, some of the leading electromagnetic effects on the thermodynamic properties of the system can be captured while neglecting the back-reaction on the geometry.
In that case, one can just consider unbroken electromagnetism in the background of a Schwarzschild BH, as we now review.

Following \cite{CPW}, we impose boundary conditions on $A$ by fixing the value of its line integral around the compactified time direction, as follows:
\begin{equation}
	\oint_{S^1} A \equiv \tilde \omega \qquad {\rm at} \qquad r=r_B \ .
\label{eq:Abc}
\end{equation}
The spherically symmetric solution to Maxwell's equations on a 
BH background, and that satisfies Eq.(\ref{eq:Abc}) reads
\begin{equation}
	A_\tau (r) = \frac{\tilde \omega}{4 \pi r_+} \left( 1 - \frac{r_+}{r_B} \right)^{-1} \left( 1 - \frac{r_+}{r} \right) \ ,
\label{eq:AtauU1}
\end{equation}
where $r_+$ should be understood as a function of $\beta$ and $r_B$ as defined through $\beta = 4 \pi r_+ \sqrt{g_{\tau \tau} (r_B)}$.

The action describing the system now includes the gravitational piece as well as the electromagnetic bulk term:
\begin{equation}
	S_{\rm EM} = \int d^4 x \ \sqrt{g} \ \frac{1}{4} g^{\mu \alpha} g^{\nu \beta} F_{\mu \nu} F_{\alpha \beta} \ .
\label{eq:EMaction}
\end{equation}
Neglecting back-reaction, the total action evaluated on the classical solution reads
\begin{equation}
	S_E (\beta, \tilde \omega, r_B) = S_g (\beta, r_B) + \frac{{\tilde \omega}^2}{2} \left( 1 - \frac{r_+}{r_B} \right)^{-1} \ ,
\label{eq:SERN}
\end{equation}
with $S_g (\beta, r_B)$ as in Eq.(\ref{eq:deltaSBH}).

\begin{figure}[h]
    \centering
    \includegraphics[scale=0.65]{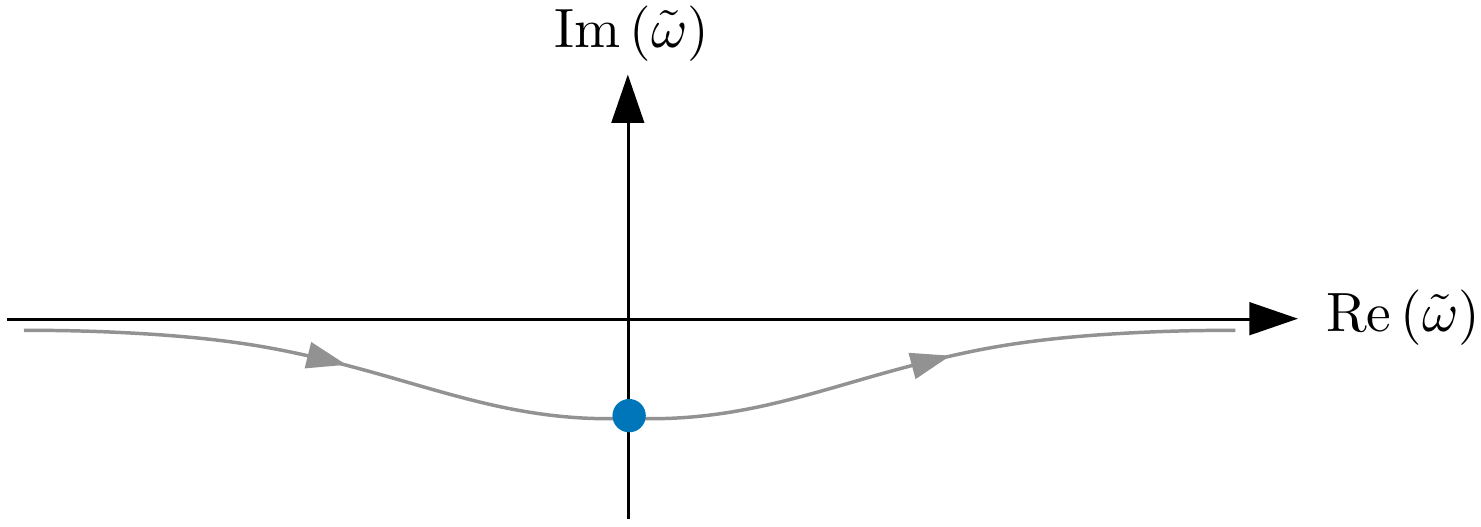}
\caption{The integrand in Eq.(\ref{eq:ZQ}) features a single saddle point that lies on the imaginary axis (blue dot).
	Within the saddle-point approximation, the contour of integration is deformed so that it passes through the saddle point (grey line), and the leading contribution
	to the integral is given by evaluating the integrand on the saddle point solution.}
\label{fig:saddleU1}
\end{figure}

We can now project the grand-canonical partition function into a sector of fixed charge by inserting the appropriate projection operator \cite{CPW}:
\begin{align}
	Z (\beta, Q, r_B) & \propto \int_{- \infty}^\infty d {\tilde \omega} \ e^{ - i Q \tilde \omega } Z(\beta, \tilde \omega, r_B) \\
				& \simeq \int_{- \infty}^\infty d {\tilde \omega} \ e^{ - \left[ i Q \tilde \omega + S_E (\beta, \tilde \omega, r_B) \right] } \ ,
\label{eq:ZQ}
\end{align}
where the last step only holds at leading order in the semiclassical expansion, and we have ignored an overall normalization factor that will be irrelevant in what follows.
The integral in Eq.(\ref{eq:ZQ}) can be evaluated in the saddle-point approximation.
The integrand has a single saddle point at a purely imaginary value of $\tilde \omega$, given by
\begin{equation}
	{\tilde \omega}_* = - i Q \left( 1 - \frac{r_+}{r_B} \right) \ .
\end{equation}
The leading order contribution to the canonical partition function is obtained by evaluating the right-hand-side of Eq.(\ref{eq:ZQ}) at the saddle point $\tilde \omega = {\tilde \omega}_*$ (see Figure~\ref{fig:saddleU1}).
This leads to an extra contribution to the free energy of the system, given by
\begin{equation}
	\Delta F (\beta, Q, r_B) \equiv F (\beta, Q, r_B) - F (\beta, r_B) \simeq \frac{Q^2}{2 \beta} \left( 1 - \frac{\beta}{4 \pi r_B} \right) \ .
\label{eq:deltaFQ}
\end{equation}
As before, we may now obtain all other thermodynamic quantities by differentiation.
For instance, the internal energy of the ensemble now includes an extra term
\begin{equation}
	\Delta U (\beta, Q, r_B) = \left( \frac{\partial (\beta \Delta F)}{\partial \beta} \right)_{Q, r_B} \simeq - \frac{Q^2}{8 \pi r_B} \ ,
\label{eq:deltaURN}
\end{equation}
which corresponds to minus the electrostatic self-energy of the system.

\begin{figure}[h]
    \centering
    \includegraphics[scale=0.65]{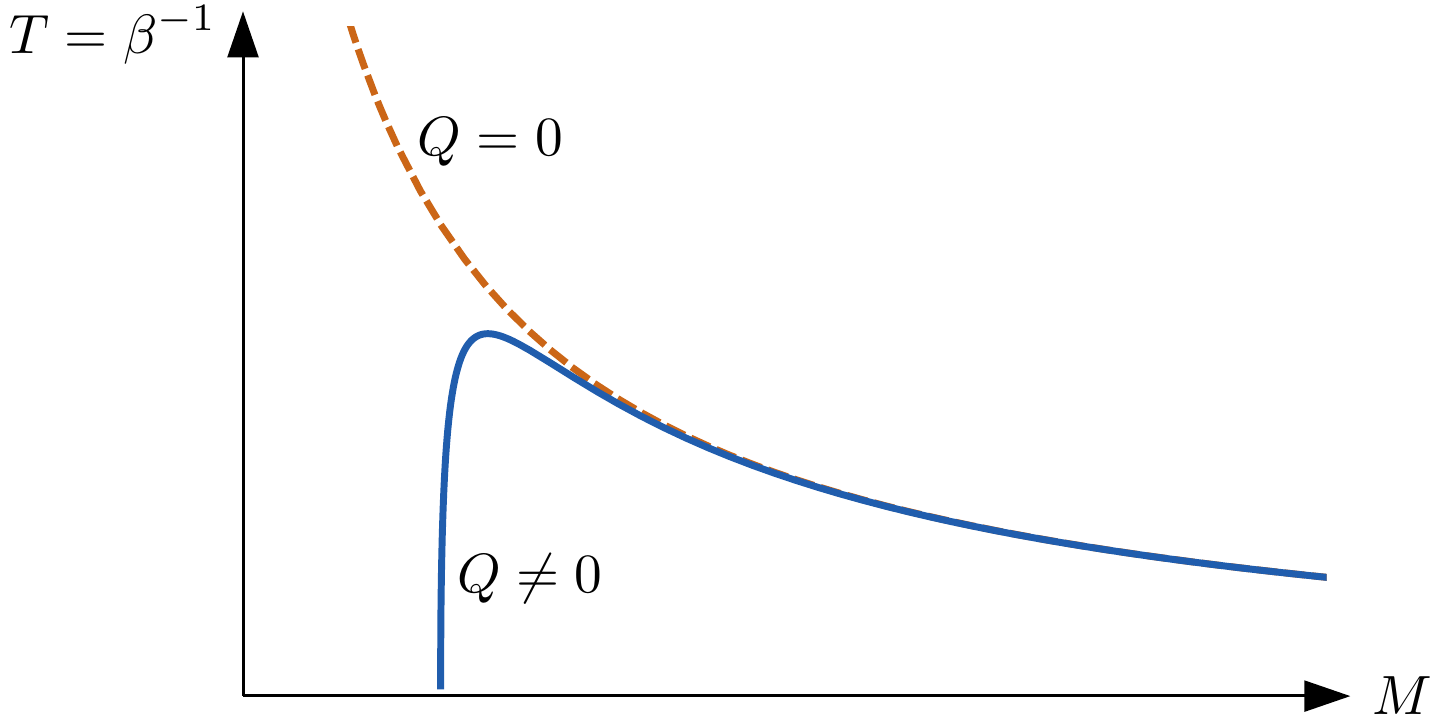}
\caption{Temperature vs mass for a Schwarzschild (dashed line) and Reissner-Nordstr\"om (solid line) BH when $r_B \rightarrow \infty$.
	A BH carrying unscreened electric charge is always colder than its uncharged counterpart.}
\label{fig:TvsM}
\end{figure}

In the limit $r_B \rightarrow \infty$, the leading charge-dependent contribution to the free energy in Eq.(\ref{eq:deltaFQ}) remains, but Eq.(\ref{eq:deltaURN}) vanishes.
Capturing the leading correction to the BH mass in this case requires including the gravitational back-reaction of the electric charge,
and therefore considering the full Reissner-Nordstr\"om solution.
Figure~\ref{fig:TvsM} shows the qualitative form of the temperature vs mass relationship for Schwarzschild and Reissner-Nordstr\"om BHs in the $r_B \rightarrow \infty$ limit:
the temperature of a charged BH is always below that of a Schwarzschild BH of the same mass.
In the region far from extremality, 
the leading correction to the mass of a BH at temperature $\beta^{-1}$, and carrying charge $Q$ reads
\begin{equation}
	\Delta M (\beta, Q) \simeq - \frac{2 \pi G Q^4}{\beta^3} \ ,
\label{eq:deltaMunscreened}
\end{equation}
and is suppressed with respect to the mass of a Schwarzschild BH at the same temperature, $\beta / (8 \pi G)$, by a factor $\sim (G Q^2 / r_+^2)^2 \ll 1$.

We can also find the expression for the classical electric field corresponding to our charged BH.
From Eq.(\ref{eq:AtauU1}) evaluated on the saddle point, we find $F_{r \tau} = \partial_r A_\tau = - i Q / (4 \pi r^2)$.
Analytically continuing this result to real time, the radial component of the electric field is given by
\begin{equation}
	E_r (r) = \frac{Q}{4 \pi r^2} \ ,
\label{eq:ErU1}
\end{equation}
as corresponds to a spherical distribution of charge. 

In the remainder of this paper, we study the effect of screened electric charge while neglecting back-reaction. Unlike for unscreened charge, a non-zero contribution to the internal energy of the system
survives in the $r_B \rightarrow \infty$ limit. As will be discussed in section~\ref{sec:discreteInfiniteBox}, this contribution is exponentially suppressed, but has the same sign as Eq.(\ref{eq:deltaMunscreened})
both in the thin- and thick-string limits.
Extending our treatment to include back-reaction is certainly a worthwhile endeavour: 
it has the potential to reveal new extremal solutions stabilised only by quantum hair,
as well as being a necessary ingredient to fully realise the $v \rightarrow 0$ limit that would interpolate between the dashed and solid lines of Figure~\ref{fig:TvsM}.

\section{Screened electric charge in an infinite-sized system}
\label{sec:discreteInfiniteBox}

In section~\ref{sec:discreteBasics} we review some basic properties of discrete gauge symmetries in a BH background in the context of infinite, asymptotically flat space, following \cite{CPW}.
In \ref{sec:thin}, we review the thin-string analysis of \cite{CPW}, and in \ref{sec:thick} we present our results for the thick-string regime.

\subsection{General considerations}
\label{sec:discreteBasics}

As advertised in section~\ref{sec:RNBH}, we study the effects of screened electric charge on the thermodynamic properties of BHs in the limit in which the backreaction on the geometry can be neglected.
The theory we consider is therefore broken electromagnetism in the background of a Schwarzschild BH, and we first work in the limit $r_B \rightarrow \infty$.
The Euclidean action 
contains the gravitational term of Eq.(\ref{eq:SESchw}), as well as the matter contribution
\begin{equation}
	S_m = \int d^4 x \ \sqrt{g} \left[ \frac{1}{4} g^{\mu \alpha} g^{\nu \beta} F_{\mu \nu} F_{\alpha \beta} +
									g^{\mu \nu} (D_\mu \phi)^\dagger D_\nu \phi + V(\phi) \right] \ ,
\label{eq:SEmatter}
\end{equation}
where $D_\mu \phi = (\partial_\mu - i g p A_\mu ) \phi$, and the scalar potential reads
\begin{equation}
	V (\phi) = \frac{\lambda}{2} \left( |\phi|^2 - \frac{v^2}{2} \right)^2 \ .
\end{equation}
As with the standard treatment of two-dimensional vortices (see e.g.~\cite{Preskill:1986kp}), we will be able to infer the existence and main properties of the relevant instanton solutions simply by considering the
qualitative features of Eq.(\ref{eq:SEmatter}).

Since all three terms in Eq.(\ref{eq:SEmatter}) are positive definite, any field configuration with finite action must be such that the contribution from each of these terms is itself finite.
The scalar potential term forces $|\phi| \rightarrow v / \sqrt{2}$ as $r \rightarrow \infty$.
This still allows for a $\tau$-dependent phase in the asymptotic behaviour of $\phi$, which without loss of generality can be written as
\begin{equation}
	\phi \rightarrow \frac{v}{\sqrt{2}} e^{\frac{i 2 \pi k \tau}{\beta}} \ , \qquad {\rm with} \qquad k \in \mathbb Z \ .
\label{eq:phiasym}
\end{equation}
The integer $k$ labels different ``vorticity'' sectors of the scalar field.
Crucially, field configurations of different $k$ cannot be smoothly deformed into one another (that is, while maintaining finite action), and so the smallest action configuration with a given $k$ is in fact a solution to the classical field equations.

With the large $r$ behaviour of the scalar field as in Eq.(\ref{eq:phiasym}), the second term in Eq.(\ref{eq:SEmatter}) being finite now requires that the $A_\tau$ piece in the $\tau$-covariant
derivative term matches the asymptotic form of $\partial_\tau \phi$, i.e.
\begin{equation}
	A_\tau \rightarrow \frac{2 \pi k}{\beta g p} \ .
\label{eq:Aasym}
\end{equation}
The one-form potential is then pure gauge at spatial infinity, which in turn ensures that $F$ vanishes and therefore the first term in Eq.(\ref{eq:SEmatter}) also remains finite.

The exact expression for the asymptotic behaviour of $\phi$ and $A_\tau$ is of course gauge dependent.
However, there is no non-singular gauge transformation that allows us to completely remove the $\tau$-dependence of the phase of $\phi$, and set $A_\tau$ to zero everywhere.
In other words, the consequences of the non-trivial asymptotic behaviour of the matter fields in Eq.(\ref{eq:phiasym}) and (\ref{eq:Aasym}) can be gauged away locally but not globally.
The best way to express this is through the line integral of $A$ along the compactified time direction at spatial infinity:
\begin{equation}
	\oint_{S^1} A = \lim_{r \rightarrow \infty} \int_0^{\beta} d \tau A_\tau = \frac{2 \pi k}{g p} \ .
\label{eq:intAthin}
\end{equation}
This expression is gauge invariant, and through Stoke's theorem is related to the surface integral of $F$ at $r \rightarrow \infty$.
Interpreting $F_{r \tau}$ as a magnetic field component in the direction perpendicular to the $r-\tau$ plane, the amount of flux going through the plane is precisely given by Eq.(\ref{eq:intAthin}),
and as we can see is quantized in units of $2 \pi / (g p)$.
The amount of magnetic flux quanta in each $k$-sector is precisely $k$ -- the vorticity of the scalar field.
Notice Eq.(\ref{eq:intAthin}) is the same as Eq.(\ref{eq:Abc}) -- the boundary condition we imposed on $A$ to obtain the partition function in a sector of fixed electric charge -- but with $\tilde \omega \rightarrow 2 \pi k / (gp)$.

\begin{figure}[h]
    \centering
    \includegraphics[scale=0.65]{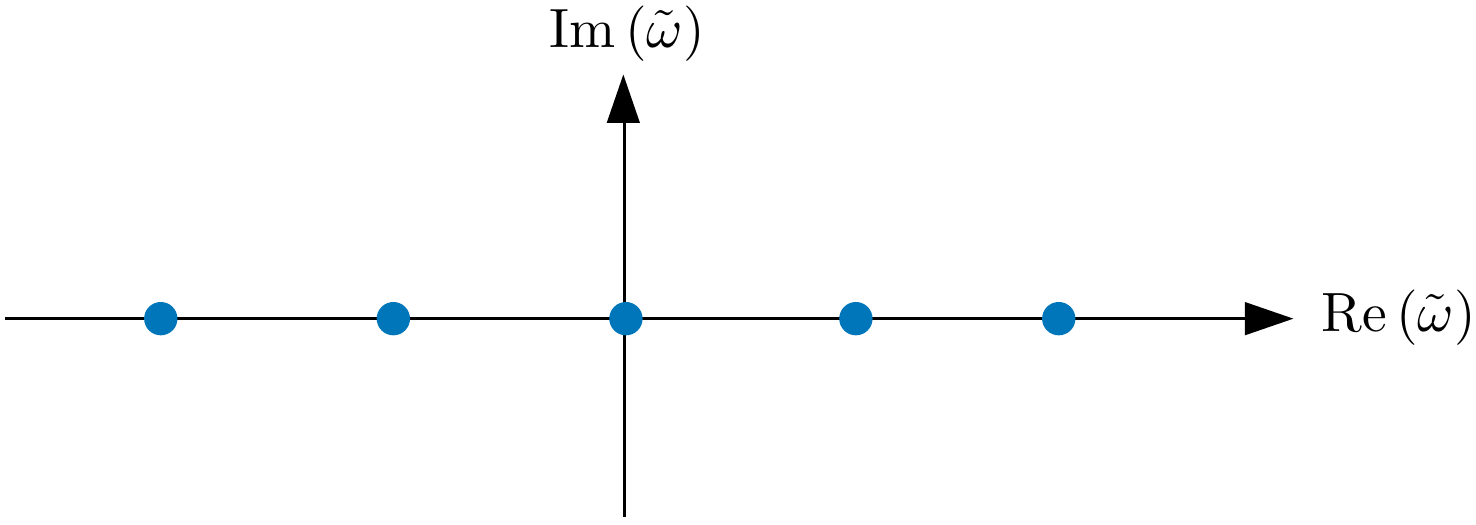}
\caption{The integrand in Eq.(\ref{eq:ZQdiscreteFull}) features an infinite number of saddle points, one for each vorticity (blue dots).
	All the saddle points lie in the real line, and are given by ${\tilde \omega}_k = 2 \pi k / (gp)$.
	Notice the difference with respect to the result for unscreened electric charge, where the only saddle point lies on the imaginary axis (see Figure~\ref{fig:saddleU1}).}
\label{fig:saddleZ}
\end{figure}

The partition function in a sector with discrete charge $Q$ should now include a discrete sum over the different vorticity sectors, as well as an integral over $\tilde \omega$, i.e.~
\begin{equation}
	Z (\beta, Q) \propto \sum_{k=-\infty}^{+\infty} \int_{-\infty}^{+\infty} d {\tilde \omega} \ e^{- i Q \tilde \omega} Z_k (\beta, \tilde \omega) \ ,
\label{eq:ZQdiscreteFull}
\end{equation}
where $Z_k(\beta, \tilde \omega)$ denotes the partition function evaluated on a sector with vorticity $k$. 
The leading order contribution to $Z(\beta, Q)$ within the semiclassical expansion can now be obtained by saddle-point evaluating the integrand on each vorticity sector,
and subsequently performing the discrete sum over $k$.
For an infinite-sized system, the value of the saddle point in each $k$-sector is just ${\tilde \omega}_k = 2 \pi k / (g p)$, as this corresponds to the only field configuration with finite action for each vorticity
(see Figure~\ref{fig:saddleZ}).
The expression for the partition function is then just given by \cite{CPW}
\begin{equation}
	Z (\beta, Q) \propto \sum_{k=-\infty}^{+\infty} e^{- i Q {\tilde \omega}_k} Z_k (\beta, {\tilde \omega}_k) \qquad {\rm with} \qquad {\tilde \omega}_k = \frac{2 \pi k}{g p} \ .
\label{eq:ZQsum}
\end{equation}
It will be useful to think of this result as due to the exponential of the integrated $\tau$-covariant derivative term in Eq.(\ref{eq:SEmatter}) acting in Eq.(\ref{eq:ZQdiscreteFull}) as a delta-function $\sim \delta (\tilde \omega - {\tilde \omega}_k)$ in the
$r_B \rightarrow \infty$ limit: when $\tilde \omega \neq {\tilde \omega}_k$, the value of the Euclidean action diverges, and the contribution to the integral in Eq.(\ref{eq:ZQdiscreteFull}) vanishes. 

The value of the matter action in Eq.(\ref{eq:SEmatter}) grows with $| k |$, and therefore the contribution from larger vorticity sectors to the sum in Eq.(\ref{eq:ZQsum}) becomes exponentially suppressed.
The leading contribution comes from the $k=0$, and $k=\pm 1$ solutions.
Normalized with respect to the partition function of a BH carrying zero charge, one finds
\begin{equation}
	\frac{Z (\beta, Q)}{Z (\beta)} \simeq 1 - 2 \left[ 1 - \cos \left( \frac{2 \pi Q}{g p} \right) \right] e^{- \Delta S_v} \ ,
\label{eq:ZQdiscrete}
\end{equation}
where $\Delta S_v$, the vortex action, corresponds to the action difference between the $k=1$ and $k=0$ sectors.
From Eq.(\ref{eq:ZQdiscrete}), we can obtain the change in free energy due to discrete charge, which is given by
\begin{equation}
	\Delta F(\beta, Q) \simeq \frac{2}{\beta} \left[ 1 - \cos \left( \frac{2 \pi Q}{g p} \right) \right] e^{- \Delta S_v} \ .
\label{eq:deltaFQdiscrete}
\end{equation}
As before, other thermodynamic properties may be obtained by differentiating Eq.(\ref{eq:deltaFQdiscrete}).
In particular, the change in internal energy of the ensemble, which in the $r_B \rightarrow \infty$ limit we may identify with the system's mass, now reads
\begin{equation}
	\Delta U (\beta, Q) \simeq - 2 \left( \frac{\partial \Delta S_v}{\partial \beta} \right)_Q \left[ 1 - \cos \left( \frac{2 \pi Q}{g p} \right) \right] e^{- \Delta S_v} \ .
\label{eq:deltaUQdiscrete}
\end{equation}
The sign of this correction will crucially depend on the $\beta$-dependence of the vortex action.

In order to evaluate $\Delta S_v$, we need to know the behaviour of $\phi$ and $A$ away from spatial infinity.
For $k=0$, $\phi \equiv v /\sqrt{2}$ and $F \equiv 0$ everywhere, leading to a vanishing matter action.
The sector of zero vorticity therefore corresponds to the vacuum state of the theory.
On the other hand, when $k \neq 0$ finiteness of the action 
demands that both $\phi$ and $A_\tau$ vanish at $r = r_+$.
An ansatz for the $k=1$ sector can then be written as
\begin{align}
	\phi (r) & = \rho(r) \frac{v}{\sqrt{2}} e^{ \frac{i 2 \pi \tau}{\beta} }
\label{eq:phiAns} \\
       A_\tau (r) & = a(r) \frac{2 \pi}{\beta g p} \ ,
\label{eq:AtauAns}
\end{align}
with $\rho(r), a(r) \rightarrow 1$ as $r \rightarrow \infty$, and $\rho(r_+) = a(r_+) = 0$.

As mentioned in the Introduction, a feature that crucially differentiates screened from unscreened hair is the absence in the former of a classical electric field.
In the $k=0$ sector, which corresponds to the vacuum state of the theory, $F \equiv 0$ everywhere, and therefore there is no classical value of the electric field for discrete charge.
A non-zero electric field is only present in the sectors with non-zero vorticity. 
The expectation value of the electric field, in the operator sense, can then be obtained by summing the contribution from all vorticity sectors \cite{CPW}.
The contribution from the $k=\pm 1$ sector dominates the sum, and the result can be written as (up to an irrelevant $\mathcal{O} (1)$ constant):
\begin{equation}
	\langle E_r (r) \rangle \sim \sin \left( \frac{2 \pi Q}{g p} \right) F_{r \tau} (r) e^{- \Delta S_v} \ ,
\end{equation}
where $F_{r \tau} (r) = \partial_r A_\tau (r)$, with $A_\tau (r)$ corresponding to the ansatz in Eq.(\ref{eq:AtauAns}).

In the following, we will find the parametric form of $\Delta S_v$ in the thin- and thick-string limits by refining our ansatz for the scalar and vector fields in these two limiting cases.
We make the simplifying assumption that $\phi$ and $A$ remain far from their asymptotic values in regions of size $\Delta r_s$ and $\Delta r_v$ from the BH horizon respectively,
and quickly reach their asymptotic behaviour outside.
Evaluating Eq.(\ref{eq:SEmatter}) on this ansatz, and minimizing the result with respect to $\Delta r_s$ and $\Delta r_v$, will provide us with the leading parametric dependence of $\Delta S_v$ on $\beta$.
Throughout, we assume that $\lambda, g p = \mathcal{O}(1)$, so that the scalar and vector masses are $m_s \sim m_v \sim v$.

\subsection{Thin-string limit}
\label{sec:thin}

In this section, we review the thin-string case discussed in \cite{CPW}, which is realised in the limit $v r_+ \gg 1$.
Neglecting back-reaction in this regime requires $G v^2 \ll 1$, and we will assume this is the case throughout.
Under the assumption (to be later justified) that $\Delta r_v, \Delta r_s \ll r_+ $, the different contributions to Eq.(\ref{eq:SEmatter}) in the sector with $k=1$ vorticity can be written, parametrically, as follows.
From the $F^2$ term, assuming $A_\tau$ has unbroken behaviour for $r -r_+ \lesssim \Delta r_v$, and is pure gauge elsewhere, we find
\begin{equation}
	S_{F^2} \sim \frac{\pi^2}{(g p)^2} \frac{r_+}{\Delta r_v} \ ,
\label{eq:thinF2}
\end{equation}
which corresponds to the magnetic self-energy of the system.
From the scalar potential term:
\begin{equation}
	S_{V} \sim \pi^2 \lambda v^4 r_+^3 \Delta r_s \ .
\label{eq:thinV}
\end{equation}
The second term in Eq.(\ref{eq:SEmatter}) contains two different pieces. When $\Delta r_v > \Delta r_s$, the $\tau$-covariant derivative term gives a contribution of the form
\begin{equation}
	S_{\tau} \sim \pi^2 (r_+ v)^2 \log \frac{\Delta r_v}{\Delta r_s} \ ,
\label{eq:thinTau}
\end{equation}
and from the $|D_r \phi|^2$ term, we have
\begin{equation}
	S_{r} \sim \pi^2 (r_+ v)^2 \left( 1 + \frac{2 \Delta r_s}{3 r_+} \right) \ .
\label{eq:thinR}
\end{equation}
The leading parametric dependence of $\Delta r_v$ and $\Delta r_s$ can be found by minimizing Eq.(\ref{eq:thinF2})-(\ref{eq:thinTau}), leading to
\begin{equation}
	\Delta r_v \sim \frac{1}{r_+ m_v^2} \ , \qquad {\rm and} \qquad \Delta r_s \sim \frac{1}{r_+ m_s^2} \ .
\end{equation}
Notice $\Delta r_v, \Delta r_s \ll r_+$ so long as $v r_+ \gg 1$, justifying our earlier assumption.

Evaluating the matter action on this solution, we find all four terms give a contribution of the same size, which parametrically may be written as
\begin{equation}
	\Delta S_{v, {\rm thin}} \sim 4 \pi r_+^2 T_s \sim \beta^2 v^2 \ ,
\label{eq:deltaSvThin}
\end{equation}
where $4 \pi r_+^2$ is the area of the BH event horizon, and $T_s \simeq \pi v^2$ corresponds to the tension of a cosmic string carrying one unit of magnetic flux.
This is as expected for the Euclidean action of an instanton solution that corresponds to the process described in the Introduction, whereby a string carrying one unit of magnetic flux envelops the BH --
namely the worldsheet area of the instanton times the string tension.

Since $\Delta S_{v, {\rm thin}} \sim (v r_+)^2 \gg 1$, by assumption in the thin-string regime, the effect due to discrete charge on the thermodynamic properties of the BH is highly exponentially suppressed.
In the limit $v r_+ \rightarrow \infty$, the correction to the free energy of the system vanishes all together, and the standard properties of a Schwarzschild BH are recovered.
The $\beta$-dependence of the vortex action in Eq.(\ref{eq:deltaSvThin}) leads to a correction to the internal energy of the system, of the form
\begin{equation}
	\Delta U (\beta, Q) \sim - \beta v^2 \left[ 1 - \cos \left( \frac{2 \pi Q}{g p} \right) \right] e^{- \Delta S_{v, {\rm thin}}} \ .
\end{equation}
Thus, the mass of a BH carrying discrete charge is smaller, albeit by an exponentially small amount, compared to that of an uncharged BH at the same temperature.
Alternatively, this translates into the charged BH being \emph{colder} than its uncharged counterpart.
Qualitatively, the effect of screened hair in this regime is the same as that of unscreened electric charge.

\subsection{Thick-string limit}
\label{sec:thick}

We now turn to the thick-string limit, realised when $v r_+ \ll 1$.
Assuming that $\Delta r_v, \Delta r_s \gg r_+$, we find the following contributions to the vortex action.
From the $F^2$ term, 
we now have
\begin{equation}
	S_{F^2} \sim \frac{\pi^2}{(g p)^2} \left( 1 + \frac{r_+}{\Delta r_v} \right) \ ,
\label{eq:thickF2}
\end{equation}
whereas from the scalar potential piece:
\begin{equation}
	S_V \sim \pi^2 r_+ \lambda v^4 \Delta r_s^3 \ .
\label{eq:thickV}
\end{equation}
As before, the $\tau$-covariant derivative term in Eq.(\ref{eq:SEmatter}) gives a non-zero contribution when $\Delta r_v > \Delta r_s$, which now reads
\begin{equation}
	S_\tau \sim \pi^2 r_+ v^2 (\Delta r_v - \Delta r_s) \ ,
\label{eq:thickTau}
\end{equation}
and the $r$-covariant derivative term contributes as
\begin{equation}
	S_r \sim \pi^2 r_+ v^2 \Delta r_s \ .
\label{eq:thickR}
\end{equation}

Minimizing Eq.(\ref{eq:thickF2}) and (\ref{eq:thickTau}) with respect to $\Delta r_v$ leads to
\begin{equation}
	\Delta r_v \sim \frac{1}{m_v} \ ,
\end{equation}
i.e.~the size of the region around the event horizon where the gauge symmetry behaves approximately as unbroken is of order $\sim m_v^{-1}$.
(The same conclusion follows by looking at Maxwell's equations in the regime $r \gg r_+$.)
To find $\Delta r_s$, we need to balance Eq.(\ref{eq:thickV}) against Eq.(\ref{eq:thickTau})-(\ref{eq:thickR}).
The exact answer will depend on the relative size of the coefficients of the $S_\tau$ and $S_r$ pieces.
If the contribution from Eq.(\ref{eq:thickTau}) is dominant, we find $\Delta r_s \sim m_s^{-1}$, whereas if the term in Eq.(\ref{eq:thickR}) were larger, then $\Delta r_s$ could be much smaller.
Determining the exact value of $\Delta r_s$ could be done numerically, by solving the coupled vector-scalar equations of motion on a Schwarzschild background.

However, the parametric form of $\Delta S_v$, including its leading $\beta$-dependence, is largely independent of $\Delta r_s$, and is parametrically given by
\begin{equation}
	\Delta S_{v, {\rm thick}} \sim \frac{\pi^2}{(g p)^2} \left( 1 + \alpha \beta v \right) \ ,
\label{eq:deltaSvThick}
\end{equation}
where $\alpha$ is a positive $\mathcal{O}(1)$ constant.
The first term in this expression is just the first term in Eq.(\ref{eq:thickF2}), and is independent of $\beta$.
The second term, which is of order $\sim v r_+ \ll 1$, captures the leading $\beta$-dependence of the vortex action, and could potentially receive contributions from all terms in Eq.(\ref{eq:thickF2})-(\ref{eq:thickR}).
Inspecting the different terms that contribute to $\Delta S_v$ in the thick-string limit reveals that, regardless of the exact size of $\Delta r_s$, the coefficient of this second term is indeed positive, and $\mathcal{O}(1)$.
Since we are only concerned with the basic qualitative effects of discrete gauge charge in this regime, this level of approximation will be good enough for our purposes.

Although the first term in Eq.(\ref{eq:deltaSvThick}), which sets the overall size of the vortex action, agrees with the result found in \cite{CPW},
\footnote{Beware of the different electromagnetic units used in \cite{CPW}. In this work, we have consistently used Lorentz-Heaviside units, where the kinetic term of an Abelian gauge theory is normalized as
in Eq.(\ref{eq:EMaction}), whereas in \cite{CPW} the Gaussian convention is implemented instead. Translating the results of \cite{CPW} to our convention requires rescaling $g^2 \rightarrow g^2 / (4 \pi)$.}
the second term, which captures the leading $\beta$ dependence, does not.
This has important consequences when it comes to determining the thermodynamic properties of the ensemble by differentiating the partition function.
Specifically, we find that the change in mass in the thick-string case reads
\begin{equation}
	\Delta U (\beta, Q) \sim - \frac{\pi^2}{(g p)^2} v \left[ 1 - \cos \left( \frac{2 \pi Q}{g p} \right) \right] e^{- \Delta S_{v, {\rm thick}}} \ ,
\end{equation}
which is negative, as in the thin-string case. 

Our result holds so long as gravitational back-reaction can be neglected, which in this limit presumably requires $G/r_+^2 \ll v r_+ \ll 1$.
On the other hand, the authors of \cite{CPW} attempt to perform their thick-string analysis in the regime in which the effect of a non-zero vev is subleading, and back-reaction effects become dominant.
In this regime, the leading order contribution to the partition function could be obtained by setting $v \equiv 0$, and would simply correspond to that of a Reissner-Nordstr\"om BH carrying unscreened charge $Q$. Further corrections could in principle be computed as an expansion in $v r_+$.
Notice, however, this is different from the approach taken in \cite{CPW}, where the expression for the canonical partition function that was obtained under the assumption $v \neq 0$ (that is, Eq.(\ref{eq:ZQdiscrete})) is still used.
The authors of \cite{CPW} then evaluate the vortex action as the difference in Euclidean actions of the Reissner-Nordstr\"om solution (subject to the boundary condition Eq.(\ref{eq:Abc}), and including back-reaction) evaluated at $\tilde \omega = 2 \pi / (gp)$ and $\tilde \omega = 0$.
The reason this procedure is unlikely to be correct, apart from the unintuitive nature of the results, is that when $v \equiv 0$ there is no notion of scalar field vorticity, certainly no need for the asymptotic behaviour
of the gauge field to behave as in Eq.(\ref{eq:Aasym}), and the expression for the partition function as given in Eq.(\ref{eq:ZQdiscrete}) is no longer valid.

We emphasize the calculation presented here is \emph{not} the correct version of the thick-string calculation done in \cite{CPW},
as our result is only valid in a different regime.
The reasons we believe the results of \cite{CPW} to be incorrect, in the regime of applicability for which they are intended, are as expressed in the previous paragraph.


The expression for the electric field in the thick-string case can be obtained by solving Maxwell's equations on a Schwarzschild background.
At distances $r \ll m_v^{-1}$, the gauge symmetry is approximately unbroken,
whereas when $r \gg m_v^{-1}$ $A_\tau (r)$ approaches its asymptotic value exponentially fast:
\begin{equation}
	\left( A_{\tau} (r) - \frac{2 \pi}{\beta g p} \right) \sim - \frac{1}{g p} \frac{e^{- m_v r}}{r} \ .
\end{equation}
In this latter regime, the expression for $F_{r \tau}$ reads, parametrically
\begin{equation}
	F_{r \tau} (r) \sim \frac{1}{g p} \left( \frac{1}{r^2} + \frac{m_v}{r} \right) e^{-m_v r}  \ ,	
\label{eq:FrtauThick}
\end{equation}
which agrees with the result found in \cite{CPW} in the appropriate limit.
The expectation value of the radial electric field as seen by an asymptotic observer is therefore just as given by Eq.(\ref{eq:Ervortex}) after substituting $F_{r \tau} (r)$ by its expression in Eq.(\ref{eq:FrtauThick}).
The result is doubly exponentially suppressed: due to the non-zero vector mass, as made explicit in Eq.(\ref{eq:FrtauThick}), as well as due to the finite vortex action.
We emphasize that the presence of this electric field in the broken phase does not violate the letter of no-hair theorems, as its origin is purely quantum.
 
\section{Screened electric charge in a finite-sized system}
\label{sec:discreteFiniteBox}

In this section, we modify the thick-string analysis of \ref{sec:thick} by considering the system to be enclosed within a finite cavity.
We take the size of the cavity to be large compared to the Compton wavelength of the vector field, i.e.~$m_v r_B \gg 1$.
Although quantitatively the deviations from the results found in section~\ref{sec:thick} are small,
the qualitative differences are significant: in a finite-sized system, a small but non-zero value of the classical electric field emerges, which now depends on the overall size of the electric charge $Q$.

We take the scalar field to have the same asymptotic behaviour as in the infinite-sized case, that is
\begin{equation}
	\phi \rightarrow \frac{v}{\sqrt{2}} e^{\frac{i 2 \pi k \tau}{\tilde \beta}} \ ,
\label{eq:phiFinite}
\end{equation}
where the period of the Euclidean time coordinate is now $\tilde \beta = \beta / \sqrt{g_{\tau \tau} (r_B)}$.
(This is different from $\beta$ by small $r_+ / r_B$ corrections, which will be irrelevant in the following.)
Our main result rests on the observation that, with Eq.(\ref{eq:phiFinite}), finiteness of the action no longer requires the asymptotic behaviour of $A_\tau$ to exactly match the phase of $\phi$.

We describe the calculation of the canonical partition function in section~\ref{sec:finiteZQ}.
Section~\ref{sec:finiteEr} is dedicated to discussing the main features of the electric field in this regime, which further clarify the results of \ref{sec:finiteZQ}.
We also comment on how our results may extrapolate to the $v \rightarrow 0$ limit, and the properties of unscreened hair recovered.

\subsection{Evaluation of the canonical partition function}
\label{sec:finiteZQ}

By solving Maxwell's equations in a finite-sized cavity, and imposing boundary conditions as in Eq.(\ref{eq:Abc}), we find a solution
for the one-form gauge field identical to that of section~\ref{sec:thick}, except within a region of size $\sim m_v^{-1} \ll r_B$ away from $r = r_B$.
In this region, close to the cavity wall, we instead find (for general scalar vorticity)
\begin{equation}
	\left( A_\tau (r) - \frac{2 \pi k}{\tilde \beta g p} \right) \sim \frac{r_B}{r} \frac{e^{-m_v (r_B - r)}}{\tilde \beta} \left( \tilde \omega - \frac{2 \pi k}{g p} \right) \ .
\label{eq:AtauFinite}
\end{equation}

With the large-$r$ behaviour of $A_\tau$ as in Eq.(\ref{eq:AtauFinite}), the value of the Euclidean action receives extra contributions from the $F^2$ and $\tau$-covariant derivative terms in Eq.(\ref{eq:SEmatter}).
To leading order, the two contributions are identical, and can be written as
\begin{equation}
	\Delta (S_{F^2} + S_\tau) \simeq \gamma \left( \tilde \omega - \frac{2 \pi k}{g p} \right)^2 \ , \qquad {\rm with} \qquad \gamma \simeq \frac{2 \pi m_v r_B^2}{\tilde \beta} \ .
\label{eq:deltaSFinite}
\end{equation}

\begin{figure}[h]
    \centering
    \includegraphics[scale=0.65]{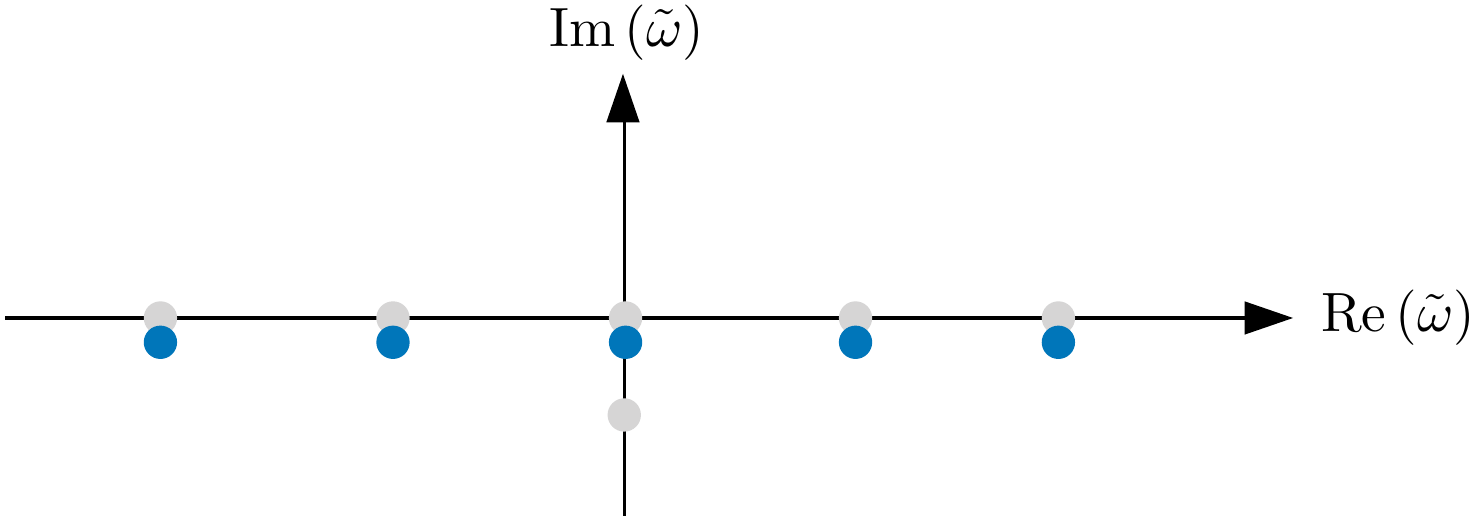}
\caption{Within a finite-sized system, the saddle points corresponding to different vorticities are displaced from the real line (blue dots).
	Although their real part is the same as in the infinite-sized case (grey dots along real axis), they also feature a small imaginary component.
	In particular, the saddle point corresponding to the $k=0$ sector moves towards the position corresponding to unscreened charge (grey dot on imaginary axis).}
\label{fig:saddleFinite}
\end{figure}

In the $r_B \rightarrow \infty$ limit, Eq.(\ref{eq:deltaSFinite}) diverges unless $\tilde \omega \rightarrow 2 \pi k / g p$, in keeping with our discussion in section~\ref{sec:discreteInfiniteBox}.
On the other hand, within a finite-sized system, the asymptotic behaviour of $A_\tau$ deviating from the scalar field vorticity no longer leads to a divergent action.
As before, we can evaluate the $\tilde \omega$ integral in Eq.(\ref{eq:ZQdiscreteFull}) in the saddle-point approximation.
With the extra contribution of Eq.(\ref{eq:deltaSFinite}), we find that the saddle point for each value of $k$ is now given by
\begin{equation}
	{\tilde \omega}_k \simeq \frac{2 \pi k}{g p} - \frac{i Q}{2 \gamma} \ .
\label{eq:saddleFinite}
\end{equation}
For all vorticity sectors, crucially including the vacuum state of the theory ($k=0$), the corresponding saddle point now features a small imaginary component.
The imaginary piece in Eq.(\ref{eq:saddleFinite}) vanishes when $r_B \rightarrow \infty$, but for finite $r_B$ it moves the position of the saddle points in Figure~\ref{fig:saddleZ} away
from the real axis.
In particular it moves the saddle point for the $k=0$ sector towards the position that corresponds to unscreened electric charge (see Figure~\ref{fig:saddleFinite}).

Saddle-point evaluating the canonical partition function, we now find
\begin{equation}
	\frac{Z (\beta, Q, r_B)}{Z(\beta, r_B)} \simeq e^{- \frac{Q^2}{4 \gamma}} \left\{ 1 - 2 \left[ 1 - \cos \left( \frac{2 \pi Q}{g p} \right) \right] e^{- \Delta S_{v, {\rm thick}}} \right\} \ ,
\label{eq:ZQFinite}
\end{equation}
with the exponential pre-factor due to the non-zero imaginary component in Eq.(\ref{eq:saddleFinite}), and $\Delta S_{v, {\rm thick}}$ as given in Eq.(\ref{eq:deltaSvThick}).
From Eq.(\ref{eq:ZQFinite}), and substituting $\gamma$ by its expression in Eq.(\ref{eq:deltaSFinite}), we find that the leading charge-dependent contribution to the free energy is now given by
\begin{equation}
	\Delta F (\beta, Q, r_B) \simeq \frac{Q^2}{8 \pi r_B^2 m_v} + \frac{2}{\beta} \left[ 1 - \cos \left( \frac{2 \pi Q}{g p} \right) \right] e^{-\Delta S_{v, {\rm thick}}} \ .
\label{eq:deltaFfinite}
\end{equation}

The free energy now contains an extra contribution, the first term in Eq.(\ref{eq:deltaFfinite}), that disappears in the limit $r_B \rightarrow \infty$.
As will become apparent in the next section, this extra term is a boundary effect, as so 
its presence cannot be associated with a change in the properties of the BH at the centre of the cavity,
which remain dominated by the second term in Eq.(\ref{eq:deltaFfinite}), and largely identical to those found in section~\ref{sec:thick}.
Its presence, however, may be potentially significant in understanding how the $v \rightarrow 0$ limit may interpolate between screened and unscreened hair.
The origin of this term becomes clearer when inspecting the properties of the electric field close to the cavity wall, to which we now turn.

\subsection{Electric field, and recovering unscreened hair}
\label{sec:finiteEr}

The large-$r$ behaviour of the gauge field found in Eq.(\ref{eq:AtauFinite}), after substituting $\tilde \omega$ by the corresponding saddle point expression,
leads to a non-zero radial electric field in the region close to the cavity wall.
The origin of this electric field is classical, as it is present in the vacuum state of the theory (that is, the $k=0$ sector).
At distances from $r = r_B$ much smaller than $m_v^{-1}$, it can be written as
\begin{equation}
	E_r (r) \simeq \frac{Q}{4 \pi r_B^2} \left( 1 - m_v (r_B - r) \right) \qquad {\rm for} \qquad r \lesssim r_B \ .
\label{Eq:ErFinite}
\end{equation}
Suggestively, this is the expression corresponding to the electric field inside a spherical shell of radius $r_B$, and thickness $m_v^{-1} \ll r_B$, at distances from the outer surface much smaller than the shell thickness.

\begin{figure}[h]
    \centering
    \includegraphics[scale=0.65]{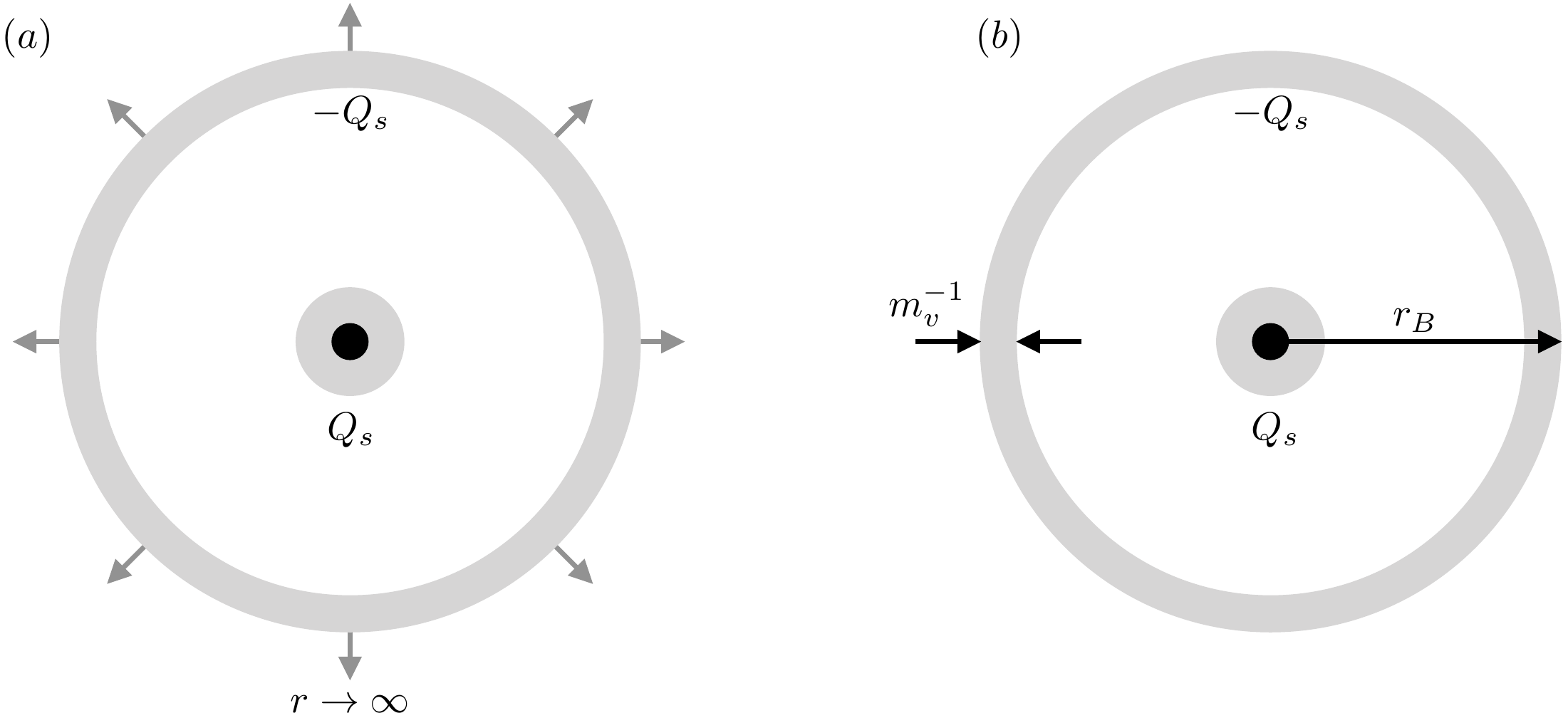}
\caption{Schematic description of how the Higgs mechanism operates, at the classical level, in the case of infinite-sized $(a)$ and finite-sized $(b)$ systems.
	A spherical distribution of charge $Q_s$ is generated at the centre of the cavity (inner grey region) that screens the charge of the BH.
	Conservation of charge requires the presence of an equal but opposite charge distribution. In a system of infinite size, this compensating charge is sent all the way to spatial infinity.
	Within a finite-sized cavity, it can only be located as far as the cavity wall, distributing itself within a region of thickness $m_v^{-1}$ (outer grey region).}
\label{fig:chargeDistribution}
\end{figure}

This result for the electric field close to the boundary is consistent with our understanding of how the process of spontaneous symmetry breaking is effective at screening charge.
At the classical level, where the notion of charge quantization plays no role, the Higgs mechanism generates a spherical distribution of charge $Q_s = - Q$ around the BH, completely screening the charge at the centre.
In order to account for charge conservation, an equal but opposite distribution, $- Q_s = Q$, needs to be present.
In an infinite-sized system, this compensating charge simply gets displaced all the way out to spatial infinity.
On the other hand, when the system is finite in extent, it can only be displaced as far as the cavity wall. The expectation is therefore that the charge would distribute itself close to the boundary,
within a region of thickness $\sim m_v^{-1}$, leading to a classical electric field in the region of non-zero charge as given by Eq.(\ref{Eq:ErFinite}).
Figure~\ref{fig:chargeDistribution} provides a schematic representation of this discussion.

The first term in Eq.(\ref{eq:deltaFfinite}) has now a clear meaning: parametrically, it corresponds to the electrostatic self-energy of the outer distribution of charge, under the assumption that
electromagnetic interactions are only effective within a range of order $m_v^{-1}$.

Our analysis also hints at how the $v \rightarrow 0$ limit may be smoothly taken, and the properties of unscreened hair recovered.
As the symmetry breaking scale gets smaller, and the Compton wavelength of the vector field grows larger, the boundary distribution of charge will get spread over a larger distance, and the inner ``screening'' charge will be localized further away from the BH.
At some point, the two will neutralize each other, and the Higgs mechanism will no longer be effective. 
The classical component of the electric field should interpolate between our result in Eq.(\ref{Eq:ErFinite}), which is only non-zero in a region of size $\sim m_v^{-1}$ close to the cavity wall, and that of the standard Reissner-Nordstr\"om solution.
On physical grounds, one would expect this transition to happen when the vector Compton wavelength becomes larger than the size of the cavity, that is $v r_B \lesssim 1$.
A more detailed analysis of the $v \rightarrow 0$ limit within a finite-sized system will be the topic of future work.

\section{Conclusions}
\label{sec:conclusions}

For any value of the $U(1)\rightarrow \mathbb{Z}_p$ symmetry breaking scale $v$, we have found that the effect of discrete charge on the mass-temperature relationship
is such that a charged BH is always colder than its uncharged counterpart. This is similar to the effect of classical (unscreened) electric hair, and resolves a puzzling feature of the results
of \cite{CPW}.   Moreover, by considering the system to be enclosed within a finite cavity, we show that the electric field present in the broken phase is not purely quantum:
a small classical component is also present.  Our analysis does not yet allow for the full $v \rightarrow 0$ limit to be realised, but it indicates that smoothly interpolating between
screened and unscreened hair requires a finite-sized system. (Alternatively, to recover the properties of $U(1)$-charged BHs within an infinite-sized system one must take the limit $v \rightarrow 0$ first.)

Our results highlight some interesting features of screened electric charge that merit further study.
In particular, back-reaction of the gauge and Higgs fields on the geometry must be taken into account for the full $v \rightarrow 0$ limit to be accessible, and the Reissner-Nordstr\"om solution recovered.
Including back-reaction is also a crucial ingredient to discover potential extremal BH solutions stabilized solely by quantum hair,
and more generally to understand the final fate of BHs carrying discrete charge \cite{Preskill:1990ty,Dvali:2007hz,Dvali:2008tq}.
Moreover, numerically solving the coupled equations of motion within a finite-sized system would provide further support for the arguments presented here, as well as new
solutions of the Euclidean Einstein-Maxwell-Higgs system relevant for the quantum properties of BHs.

Finally, we stress that the results presented here are valid when the discrete gauge symmetry is realised in the context of an Abelian Higgs model.
Other realisations of a $\mathbb{Z}_p$ gauge symmetry are nonetheless possible \cite{Banks:2010zn}, and studying the properties of charged BHs in alternative UV-completions provides a promising direction for future work.

\section*{Acknowledgments}
I am grateful to John March-Russell for many useful discussions, as well as to Mark Bowick and Nathaniel Craig for their comments on earlier versions of the manuscript.
The research of IGG is funded by the Gordon and Betty Moore Foundation through Grant GBMF7392. 
I also acknowledge financial support from Merton College, University of Oxford, through a Junior Research Fellowship, during which the majority of this work was done.
Research at KITP is supported in part by the National Science Foundation under Grant No.~NSF PHY-1748958.

\bibliography{discrete_refs}

\end{document}